\RequirePackage{amsthm}%

\documentclass[sn-mathphys-num]{sn-jnl}

\usepackage{graphicx}%
\usepackage{multirow}%
\usepackage{amsmath,amssymb,amsfonts}%
\usepackage{amsthm}%
\usepackage{mathrsfs}%
\usepackage[title]{appendix}%
\usepackage{xcolor}%
\usepackage{textcomp}%
\usepackage{manyfoot}%
\usepackage{booktabs}%
\usepackage{algorithm}%
\usepackage{algorithmicx}%
\usepackage{algpseudocode}%
\usepackage{listings}%

\theoremstyle{thmstyleone}%
\theoremstyle{thmstyletwo}%

\theoremstyle{thmstylethree}%

\begin{document}

\title[Article Title]{Design and commissioning of the miniBELEN neutron counter for the study of ($\alpha$,n) reactions}


\author*[1]{\fnm{Nil} \sur{Mont-Geli}}\email{nil.mont@upc.edu}
\author*[2]{\fnm{Ariel} \sur{Tarifeño-Saldivia}}\email{atarisal@ific.uv.es}
\author[1]{\fnm{Max} \sur{Pallàs}}
\author[5]{\fnm{Ángel } \sur{Perea}}
\author[3]{\fnm{Luis Mario} \sur{Fraile}}
\author[1]{\fnm{Guillem} \sur{Cortés}}
\author[4]{\fnm{Sílvia} \sur{Viñals}}
\author[2]{\fnm{José Luis} \sur{Tain}}
\author[4]{\fnm{Gastón} \sur{Garcia}}
\author[1]{\fnm{Daniel} \sur{Soler}}

\author[2]{\fnm{Odette} \sur{Alonso-Sañudo}}
\author[5]{\fnm{María José G.} \sur{Borge}}
\author[3,5]{\fnm{José Antonio} \sur{Briz}}
\author[1]{\fnm{Francisco} \sur{Calviño}}
\author[6]{\fnm{Daniel} \sur{Cano-Ott}}
\author[1]{\fnm{Alfredo} \sur{De Blas}}
\author[7,8]{\fnm{Begoña} \sur{Fernández}}
\author[1]{\fnm{Roger} \sur{Garcia}}
\author[4,5]{\fnm{Vicente} \sur{Garcia Tavora}}
\author[6]{\fnm{Enrique M.} \sur{González-Romero}}
\author[7,8]{\fnm{Carlos} \sur{Guerrero}}
\author[3]{\fnm{Andrés} \sur{Illana}}
\author[6]{\fnm{Trino} \sur{Martínez}}
\author[3]{\fnm{Víctor} \sur{Martínez-Nouvilas}}
\author[6]{\fnm{Emilio} \sur{Mendoza}}
\author[2]{\fnm{Enrique} \sur{Nácher}}
\author[5]{\fnm{Amanda} \sur{Nerio Aguirre}}
\author[5]{\fnm{Julio} \sur{Plaza}}
\author[5]{\fnm{Olof} \sur{Tengblad}}
\author[3]{\fnm{José Manuel} \sur{Udías}}

\affil[1]{\orgdiv{Institut de Tècniques Energètiques (INTE)}, \orgname{Universitat Politècnica de Catalunya (UPC)}, \orgaddress{\city{Barcelona}, \postcode{08028}, \country{Spain}}}
\affil[2]{\orgdiv{Instituto de Física Corpuscular (IFIC)}, \orgname{CSIC - Univ. Valencia}, \orgaddress{\city{Paterna}, \postcode{46071}, \country{Spain}}}
\affil[3]{\orgdiv{Grupo de Física Nuclear (GFN), EMFTEL and IPARCOS}, \orgname{Universidad Complutense de Madrid (UCM)}, \orgaddress{\city{Madrid}, \postcode{28040},  \country{Spain}}}
\affil[4]{\orgdiv{Centro de Micro-Análisis de Materiales (CMAM)}, \orgname{Universidad Autónoma de Madrid (UAM)}, \orgaddress{\city{Madrid}, \postcode{28049}, \country{Spain}}}
\affil[5]{\orgdiv{Instituto de Estructura de la Materia (IEM)}, \orgname{CSIC}, \orgaddress{\city{Madrid}, \postcode{28006}, \country{Spain}}}
\affil[6]{\orgdiv{Centro de Investigaciones Energéticas, Medioambientales y Tecnológicas (CIEMAT)}, \orgaddress{\city{Madrid}, \postcode{28040}, \country{Spain}}}
\affil[7]{\orgdiv{Departamento de Física Atómica, Molecular y Nuclear}, \orgname{Universidad de Sevilla (US)}, \orgaddress{\city{Sevilla}, \postcode{41012}, \country{Spain}}}
\affil[8]{\orgdiv{Centro Nacional de Aceleradores (CNA)}, \orgname{(Universidad de Sevilla (US) - Junta de Andalucía - CSIC}, \orgaddress{\city{Sevilla}, \postcode{41092}, \country{Spain}}}


\abstract{The miniBELEN detector is a moderated neutron counter based on the use of $^3$He tubes and high-density polyethylene as moderator. The detector has been designed to have a neutron detection efficiency that is nearly independent from the initial neutron energy for ($\alpha$,n) reactions with alpha-particle energies up to 15 MeV. In order to achieve that, an innovative design methodology based on the use of cadmium neutron filters has been used. Monte Carlo calculations of the detector response have been validated using a $^{252}$Cf fission source. Measurements of the relatively well-known ($\alpha$,n) thick-target yields on aluminum have been found to be consistent with data from previous works. A method has been developed to correlate the counting rate ring ratios with the mean neutron energy and validated using previous time-of-flight measurements of the aluminum ($\alpha$,n) spectra and with a $^{252}$Cf fission source.\\

\textit{Submitted to The European Physical Journal - Plus}
}

\keywords{$(\alpha,n)$ reactions, reactions yields, neutron detector design, Monte Carlo simulation, neutron multiplicity counting, detector efficiency}



\maketitle

\section{Introduction}\label{sec:Intro}

The production of neutrons through reactions induced by $\alpha$ particles plays a crucial role in many different fields \cite{westerdale2022alpha,Junghans2024alpha,cano2024white}. Specifically, in underground physics, ($\alpha,n$) reactions are one of the main sources of the neutron-induced background in underground facilities, which is typically a key issue for experiments carried out there \cite{kudryavtsev2008neutron,KUDRYAVTSEV2020164095}. Moreover, in nuclear astrophysics, ($\alpha,n$) reactions are the major source of neutrons for the slow neutron-capture nucleosynthesis process (the s-process) \cite{RevModPhys.83.157} and are involved in the weak rapid neutron-capture nucleosynthesis process (named the weak r-process, also known as the $\alpha$-process) \cite{Bliss_2017}. Other fields of interest include nuclear technologies such as fission \cite{BEDENKO2019189} and fusion reactors \cite{Cerjan_2018}, non-destructive assays for non-proliferation and spent fuel management applications \cite{broughton2021sensitivity,romano2020alpha}, and the production of radioactive isotopes for medical applications \cite{QaimSpahnScholtenNeumaier+2016+601+624}.

All of these applications are based on accurate and precise experimental data. However, most of the available data on ($\alpha,n$) reactions was measured decades ago, is incomplete and/or present large discrepancies that are not compatible with the declared uncertainties. Furthermore, in many cases, the uncertainties on the cross sections and neutron energy spectra are rather large. To address such problems, new measurements are required \cite{westerdale2022alpha}. To that end, the Measurement of Alpha Neutron Yields (MANY) collaboration was formed.

MANY is a coordinated effort involving several institutions aiming to carry out measurements of ($\alpha,n$) reaction cross-sections, production yields and neutron energy spectra. The project relies on the use of the currently existing infrastructure in Spain, in particular the $\alpha$-beams produced by the accelerators at CMAM (Madrid) \cite{redondo2021current} and (Sevilla) \cite{gomez2021research,MILLANCALLADO2024111464}; and on the use of neutron detection systems such as miniBELEN, the 4$\pi$ neutron counter described in this paper, and MONSTER, which is a time-of-flight neutron spectrometer based on the use of BC501/EJ301 liquid scintillation modules \cite{garcia2012monster}. Both systems are complemented by $\gamma$-spectroscopy measurements using the GARY array of fast LaBr$_3$(Ce) scintillation detectors \cite{vedia2017performance} with angular resolution capabilities and germanium detectors.

Most of the currently available experimental data on ($\alpha,n$) reaction cross sections and production yields have been measured using activation techniques \cite{ROUGHTON1983341}, time-of-flight spectroscopy with scintillation detectors \cite{JACOBS1983541} and direct neutron counting \cite{stelson1964cross, bair1979neutron, PhysRevC.18.1566, westerdale2022alpha}. The latter typically involves the use of moderated neutron counters. Such detection systems are based on the use of an array of thermal neutron proportional counters embedded in a block of moderating material \cite{knoll2010}.

In moderated neutron counters, when the energy spectrum of the $\alpha$-induced neutrons is unknown, such as in most ($\alpha$,n) reactions, the proper determination of the neutron production yields requires a detection efficiency nearly independent from the initial neutron energy, namely a flat efficiency. Efforts to develop such type of detectors have been carried out for decades. Of particular relevance has been the graphite sphere detector developed by R. L. Macklin in 1957 \cite{MACKLIN1957335}. Consisting on an array eight BF$_3$ proportional counters embedded in a sphere of graphite moderator, the Macklin detector was used in many of the first experimental studies on ($\alpha$,n) reactions \cite{stelson1964cross, bair1979neutron}. In more recent years new setups appeared such as  FED \cite{utsunomiya2017direct}, BRIKEN \cite{tarifeno2017conceptual}, HeBGB \cite{Brandenburg_2022} and ELIGANT-TN \cite{clisu2023cross} appeared. In section \ref{sec:detDesing} a more comprehensive review of the characteristics of these and other detectors is provided.

miniBELEN is a modular moderated neutron counter. It builds on the experience gained from previous experiments involving other detectors of the BELEN-type such as BELEN-20 \cite{gomez2011first}, BELEN-30, BELEN-48 \cite{agramunt2014new} and BRIKEN \cite{tarifeno2017conceptual}. All of these detectors are based on the use of High-Density PolyEthylene (HDPE) as neutron moderator and $^3$He-filled proportional counters as thermal neutron counters. In such type of instruments, thermal neutrons are converted through the $^{3}$He(n,p)$^{3}$H reaction into proton-tritium pairs which generate electronic pulses whose amplitude is proportional to the deposited energy. Modularity implies that the HDPE moderator is formed by several smaller blocks, which can be reassembled in different ways to provide the detector with different types of response. In practice, it means that we have various detectors for the prize of one. A modular detector also implies easiness to introduce design improvements and modifications. Moreover, it is easier to transport a modular detector than a compact one. 

An innovative design methodology based on the use of cadmium filters to weight the contribution of each neutron counter to the total response of the detector, the so-called composition method, has been used to achieve flat efficiency. This method allowed to overcome the limitations that the modular structure imposed on the possible positions of the counters within the moderator, the optimization of which is the most traditional approach to this problem \cite{falahat20133he,tarifeno2017conceptual,Brandenburg_2022}. Additionally, the method allows for a reduction in the computation time required to optimize the design.

This work is organized as follows. In the following section, we will discuss the main design requirements for the detector. The design process is discussed in section \ref{sec:detDesing} and the definitive implementation of the detector and the experimental characterization using neutron sources in section \ref{sec:Implementation}. Finally, results from the commissioning measurement through the $^{27}$Al($\alpha,n$)$^{30}$P are presented in section \ref{sec:comissioning} and some concluding remarks are given in section \ref{sec:Conclusions}.

\section{Design requirements}\label{sec:Requirements}
In most of the mentioned applications of ($\alpha$,n) reactions, in particular in underground physics and nuclear technology, the alpha particles involved come from decaying radioisotopes and present energies ranging from a few keV up to several MeV. Some astrophysical and medical applications are the most notable exception to this pattern \cite{westerdale2022alpha,QaimSpahnScholtenNeumaier+2016+601+624}. Therefore, a neutron detector with a nearly flat efficiency up to several MeV would be suitable for many different experiments involving ($\alpha,n$) reactions. The optimal value of the detection efficiency depends on each reaction. Hence, a modular structure that can be relatively easily upgraded according to the needs of each experiment implies a clear advantage. It must be kept in mind that the value of the efficiency will be limited by the number of proportional counters available and the flatness requirements. Finally, because of the non-isotropic properties of the $\alpha$-induced neutron fluxes, a nearly 4$\pi$ solid-angle coverage geometry is typically used in such types of detectors. In summary, the design requirements for miniBELEN are:
\begin{itemize}
    \item A nearly flat neutron detection efficiency up to approximately 10 MeV.
    \item A modular HDPE moderator.
    \item A nearly 4$\pi$ solid-angle coverage.
\end{itemize}

\section{Detector design}\label{sec:detDesing}
\begin{figure}[h]
\centering
    \includegraphics[width=0.8\textwidth]{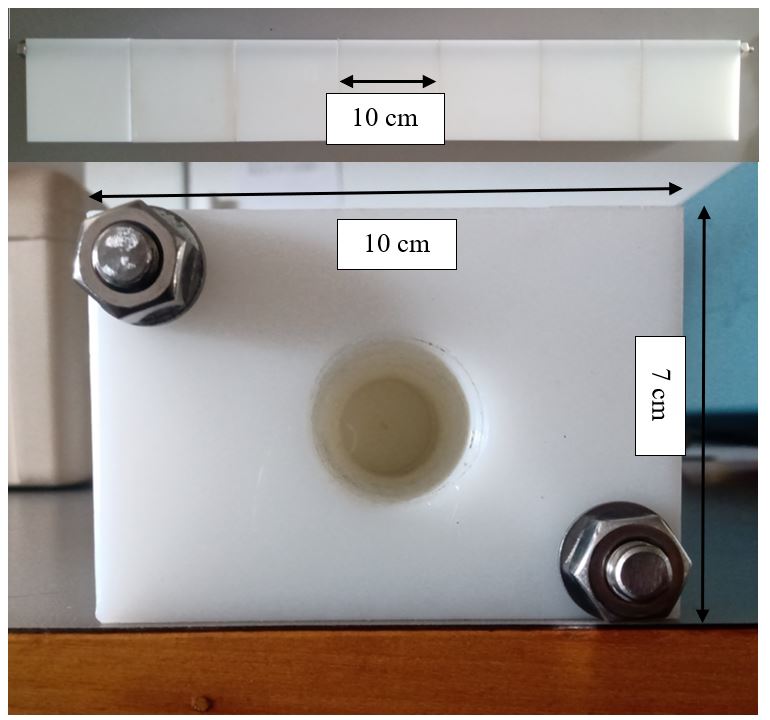}
    \caption{Upper panel: HDPE block (7x10x70 cm$^3$) used in miniBELEN. Each of these blocks is made of seven smaller pieces (7x10x10 cm$^3$) which are assembled using two stainless-steel rods. The black arrow shows one of these small blocks. Lower panel: front view of one of the HDPE blocks. The $^3$He-filled neutron counters are embedded within the central hole. The stainless-steel rods can also been seen (top left and bottom right corners).}\label{fig:HDPEblocks}
\end{figure}

The miniBELEN detector consists on an array of $^3$He-filled proportional counters sensible to thermal neutrons which are embedded in a modular moderator made of high-density polyethylene (HDPE). The design of the detector has been based on the use of readily available hardware from previous experiments that involved other detectors of the BELEN type. The basic component is a set of 7x10x70 cm$^3$ HDPE blocks with a central hole inside which the neutron counters are embedded. Each of these blocks is made of seven smaller 7x10x10 cm$^3$ pieces which are assembled using two stainless steel rods as shown in figure \ref{fig:HDPEblocks}. The density of HDPE was measured to be 0.949(2) g/cm$^3$ so that in the calculations 0.95 g/cm$^3$ was used. The use of such blocks limits the positioning of the proportional counters within the moderator, which is a crucial issue for the achievement of a nearly flat detection efficiency. 

In miniBELEN, cylindrical $^3$He-filled proportional counters manufactured by LND \cite{LND} are used to detect neutrons passing through the HDPE moderator. The counter walls are made of stainless steel. The cylinder diameter is 2.54 cm and the length is 63.72 cm. When we started to design the detector, different models with different nominal gas pressures were available. For the design calculations, a nominal 10 atm pressure (LND model 252266) was assumed. 

\subsection{Description of the simulation setup}\label{sec:simSETUP}
Extensive Monte Carlo calculations are carried out to optimize the design of the miniBELEN detector. The calculations are carried out using \textit{Particle Counter} \cite{ParticleCounter}, which is an application for Monte Carlo simulations of radiation detectors based on the Geant4 toolkit \cite{AGOSTINELLI2003250,AllisonGeant4_2006,ALLISON2016186}. In particular, \textit{Particle Counter} uses Geant4 versions 10.01 (patch 3).The simulation included the HDPE blocks, the stainless steel rods and the $^3$He proportional counters. A point-like and isotropic neutron source located at the center of the detector was assumed. 

\subsection{Figures of merit: average efficiency and flatness}\label{sec:secParMer}
Calculations are carried out using a set of discrete neutron energies, $E_i$ = \{0.0001, 0.001, 0.01, 0.1, 0.5, 1.0, 1.5, 2.0, 2.5, 3.0, ..., 9.5, 10.0\} MeV. For each energy value, the detection efficiency $\varepsilon(E_i)$ is defined as the quotient between the number of detected events and the number of primary neutrons.  Two figures of merit are defined for the optimization of the design: the \textit{average efficiency} ($\varepsilon_{av}$) and the \textit{flatness parameter} ($F$). Both are defined within a certain energy range, from 0.0001 MeV up to $E_{Max}$.
\begin{equation}
  \varepsilon_{av} (E_{Max}) = \frac{1}{N}\sum_{i = 0}^{E_{Max}}\varepsilon(E_i)
\end{equation}
\begin{equation}
  F (E_{Max}) = \frac{max[\varepsilon(E_j)]}{min[\varepsilon(E_j)]}
\end{equation}
With $E_j$ ranging from $E_0$ = 0.0001 MeV to $E_{Max}$. 

\subsection{Design methodology: the composition method}
As it has been mentioned the positioning of the proportional counters within the moderator is crucial to achieve a flat efficiency. The larger the distance between the neutron source and the proportional counter, the larger the detection efficiency for high-energy neutrons. However, increasing the moderation distance reduces the detection efficiency for low-energy neutrons. The traditional approach to designing moderated neutron counters with flat efficiencies is to organize the proportional counters in groups defined by a common distance to the neutron source (the so-called \textit{rings} of counters), which is usually located at the center of the detector. Then the design problem becomes the problem of optimizing the radii of each ring of counters \cite{falahat20133he,tarifeno2017conceptual,Brandenburg_2022}. In miniBELEN, however, the modular structure of the moderator introduces a limitation on the possible positions of the neutron counters within the moderator. Consequently, an alternative approach, the so-called composition method, has been used. This novel methodology is based on fixing the position of the neutron counters and optimizing each ring of counters to maximize the total detection efficiency. To that end, the so-called composition functions are defined in the following way (although false, it is assumed that the ring efficiencies are independent magnitudes), 

\begin{equation}
  \varepsilon (E_i) = \sum_{j = 1}^{nr}\varepsilon_{j}(E_i)f_j(E_i)
\end{equation}

Being $\varepsilon (E_i)$ the total neutron detection efficiency, $\varepsilon_{j}(E_i)$ the detection efficiency of ring $j$, $nr$ the total number of rings, $E_i$ the original neutron energy, and $f_j(E_i)$ the composition functions, taking values between 0 and 1. Therefore, the design problem becomes a problem of optimizing the composition functions in order to obtain a flat neutron detection efficiency. The use of such a methodology greatly reduces the computation time, increasing the efficiency of the optimization process. The most convenient approach is to use composition functions that are independent of the initial neutron energy $E_i$. 

\subsection{Optimization of miniBELEN}
Different configurations were created by assembling the HDPE blocks in figure \ref{fig:HDPEblocks} in different ways. These configurations were created with the aim of preserving the compactness and symmetry of the setup.  For each configuration the neutron detection efficiency was calculated using \textit{Particle Counter}. The composition method was then applied to find the optimal composition functions, which were defined as those values that minimize the flatness parameter and present a sufficiently high neutron detection efficiency (above 2\%).

\begin{figure}[!ht]
    \centering
    \includegraphics[width=0.6\textwidth]{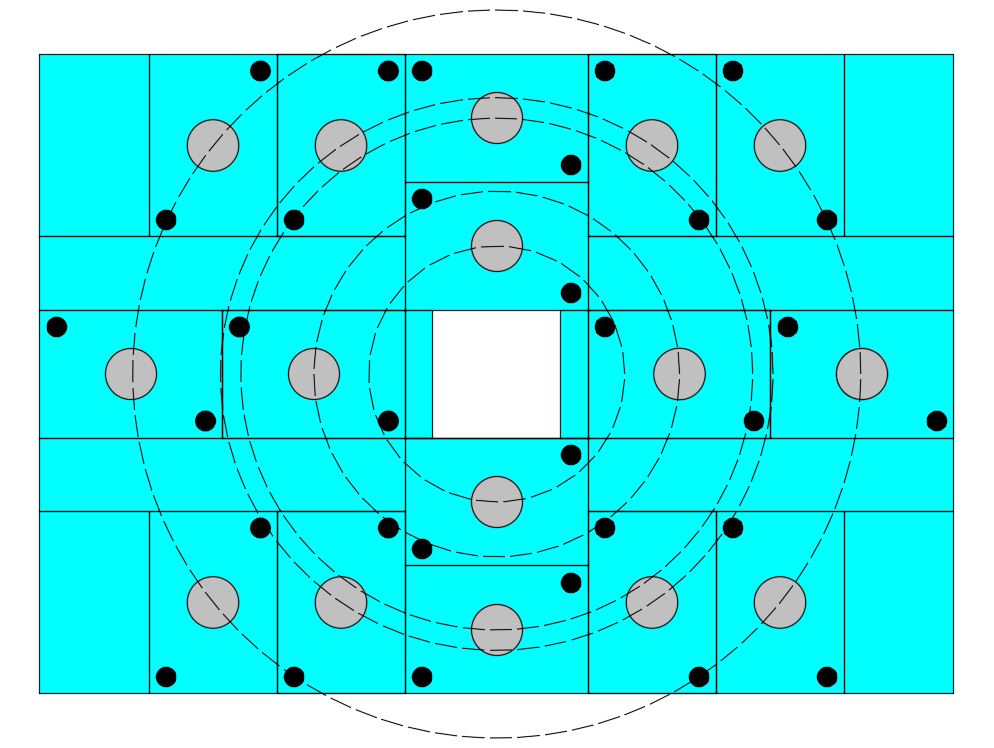}
    \caption{Selected configuration for the optimization of the composition functions (miniBELEN-X). Light-blue: HDPE. Black: stainless-steel rods. Grey: proportional counters. White: 7x7 cm$^2$ central hole required to hold the beam-pipe and gamma detectors (see text for more details). Each ring is shown dashed circles. This is the view of the detector looking from the end of the beam-line.}
    \label{fig:HDPEconfA}
\end{figure}

\begin{figure}[!ht]
    \centering
    \includegraphics[width=0.8\textwidth]{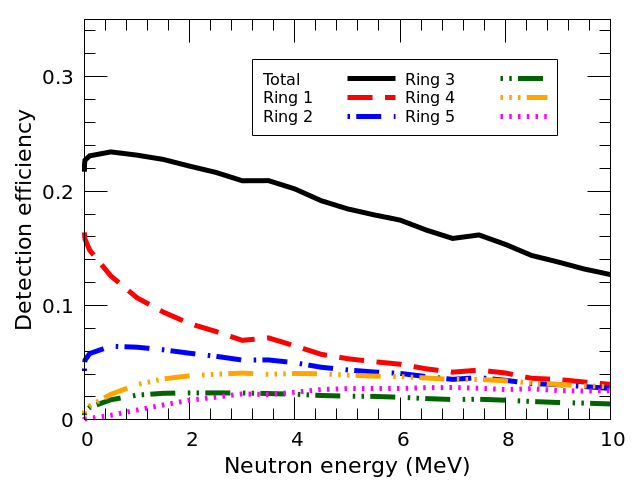}
    \caption{Neutron detection efficiency of miniBELEN-X calculated using \textit{Particle Counter}. The efficiency have been calculated using the set of neutron energies $E_i$ from section \ref{sec:secParMer}. Linear interpolations are used between the data points. The y-axis error-bars are smaller than width of the data lines. Black solid: total efficiency. Red dashed: ring 1 (inner) efficiency. Blue dash-dotted (1 dot): ring 2 efficiency. Green dash-dotted (2 dots): ring 3 efficiency. Orange dash-dotted (3 dots): ring 4 efficiency. Magenta dotted: ring 5 (outer) efficiency.}
    \label{fig:noComEffA}
\end{figure}

Off all the configurations studied, the one shown in figure \ref{fig:HDPEconfA} is the most suitable for achieving a flat efficiency (since now, miniBELEN-X). It consists of sixteen neutron counters that are distributed in 5 rings. As can be seen, the basic components are the HDPE blocks in figure \ref{fig:HDPEblocks}. Additional HDPE blocks were used to complete the detector. A 7x7 cm$^2$ central hole was considered in order to hold the beam pipe and gamma detectors (the use of high-purity germanium and LaBr$_3$(Ce) scintillation detectors is planned). Figure \ref{fig:noComEffA} shows the calculated neutron efficiency for this configuration and it can be clearly seen that it is strongly dependent on the neutron energy. 

\clearpage

\begin{figure}[!ht]
    \centering
    \includegraphics[width=1\textwidth]{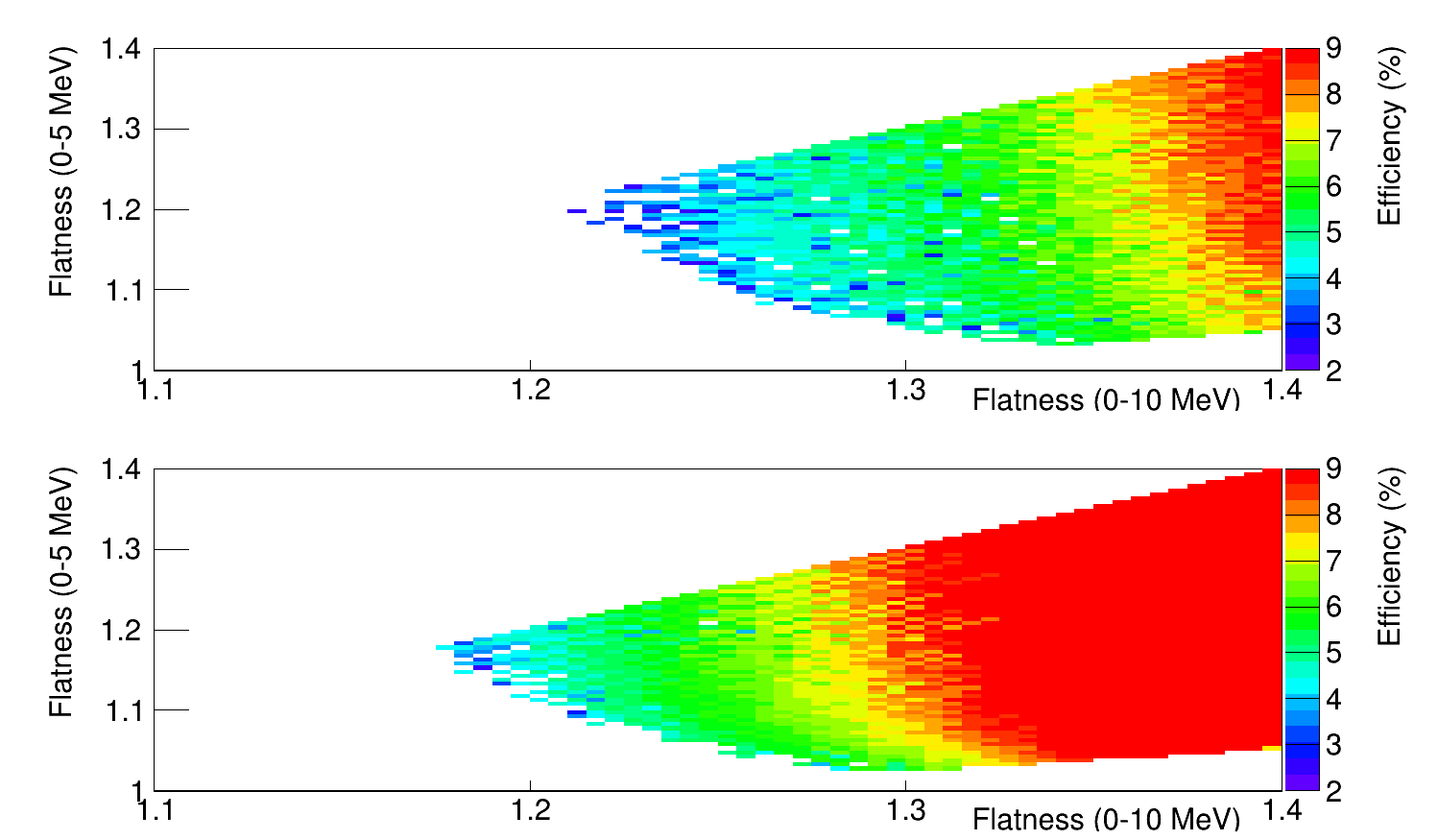}
    \caption{Average efficiencies up to 10 MeV and flatness values $F(5)$ and $F(10)$ obtained from different combinations of the composition functions. Solutions with average efficiencies below 2\% are not shown. The color scale saturates at efficiency values above 9\%. Upper panel: configuration miniBELEN-X. Lower panel: configuration miniBELEN-X + 4 cm thickness reflectors.}
    \label{fig:composition}
\end{figure}

For all possible values of the composition functions, the average efficiency up to 10 MeV and the flatness parameters $F(5)$ and $F(10)$ are calculated. Results are shown in figure \ref{fig:composition} (up), where it can be clearly seen that $F(5)$ and $F(10)$ are anticorrelated in the low-value region (the region of interest). It is also interesting that lower flatness values also imply lower efficiencies.

\begin{figure}[!ht]
    \centering
    \includegraphics[width=0.6\textwidth]{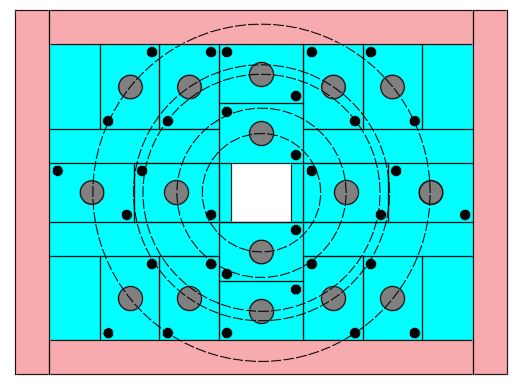}
    \caption{Reflectors (red) are HDPE which are added beyond the outer ring in order to increase the efficiency of miniBELEN without sacrificing flatness and adding new counters (see text for more details). In this example the thickness of the reflectors has been set to 4 cm. This is the rear view of the detector, looking from the end of the beam-line.}
    \label{fig:refExample}
\end{figure}

In order to increase the total efficiency without sacrificing the flatness, two complementary approaches can be used. The first one consists of the use of the so-called reflectors. These are HDPE blocks of the same type that are added beyond the outer ring (see figure \ref{fig:refExample}). Some of the higher-energy neutrons are back-scattered (reflected) so that the efficiency of the outer rings is increased. The second approach consists of adding counters in the appropriate rings and recalculating the composition functions. The consequence is that for the same values of the flatness, larger efficiencies are achieved (figure \ref{fig:composition}, down).

\clearpage

\subsection{Final configurations}
\begin{figure}[!ht]
    \centering
    \includegraphics[width=1\textwidth]{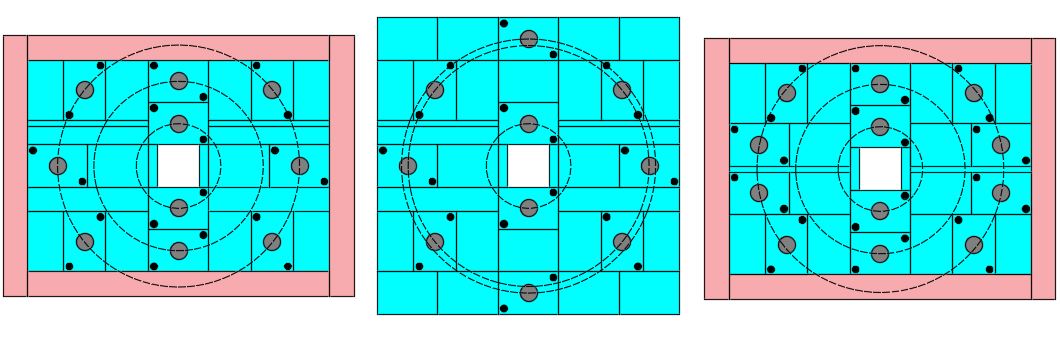}
    \caption{(From left to right) Configurations miniBELEN-10A, miniBELEN-10B and miniBELEN-12. Light-blue: HDPE. Black: stainless-steel rods. Grey: proportional counters. White: 7x7 cm$^2$ central hole required to hold the beam-pipe and gamma detectors (see text for more details). Each ring is highlighted using dashed circles. This is the rear view of the detector, looking from the end of the beam-line.}
    \label{fig:confDesigns}
\end{figure}

Three different configurations of the miniBELEN detector have been selected (a configuration is defined as any combination of values of the composition functions). The structure of the HDPE modular moderator of each one is shown in figure \ref{fig:confDesigns}. Since for all configurations the optimal composition functions for rings 2 (two counters) and 4 (four counters) in the configuration shown in figures \ref{fig:noComEffA} and \ref{fig:refExample} were found to be close to zero, it is reasonable to simply remove the affected counters and replace the HDPE blocks with holes by plain ones. Consequently, in the final configurations, ten neutron counters (instead of the initial sixteen) are distributed in a three-ring geometry (instead of the initial five). The size of the reflectors (50 cm in length and 4 cm thickness) has been optimized to achieve maximum efficiency using the minimum amount of material. It is interesting to note that each configuration is achieved by almost assembling in different ways the same HDPE blocks.

\begin{figure}[!ht]
    \centering
    \includegraphics[width=1\textwidth]{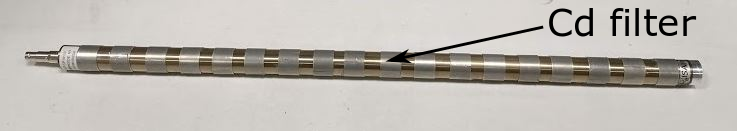}
    \caption{$^3$He-filled proportional counters are partially covered with 18 cadmium filters (2 cm length and 0.5 mm thickness) in order to physically implement the effect of the composition functions.}
    \label{fig:cdfilters}
\end{figure}

The composition functions are physically implemented by partially covering the active region of the $^3$He counters with cadmium filters (0.5 mm thickness and 2 cm length, see figure \ref{fig:cdfilters}) so that neutrons of any energy moderated below 0.5 eV are absorbed. Consequently, the effect of the cadmium filters is approximately independent of the initial neutron energy. The total length $x$ to be covered with cadmium filters in order to reproduce the effect of the composition function $f_i$ is calculated as follows:

\begin{equation}
  x = L(1-f_i)
\end{equation}

Being $L$ = 60 cm the active length of the proportional counters in miniBELEN.

\clearpage

The optimal composition functions and the number of cadmium filters required to implement them are shown in table \ref{tab:compoFunVal}. It should be kept in mind that in miniBELEN-10B and miniBELEN-12 the values of the composition functions have been adjusted to make them physically implementable using the 2 cm-long cadmium filters described in figure \ref{fig:cdfilters}.

\begin{table}[h]
    \caption{Composition functions $f_i$ and number of cadmium filters per counter $n_{i}$ in ring $i$ for each configuration of miniBELEN (see text for more details).}\label{tab:compoFunVal}
    \begin{tabular*}{\textwidth}{@{\extracolsep\fill}lcccccc}
    \toprule%
    Configuration & $f_1$ & $f_2$ & $f_3$ & $n_1$ & $n_2$ & $n_3$ \\
    \midrule
     miniBELEN-10A & 0.4 & 0.8 & 1 & 18 & 6 & 0\\
     miniBELEN-10B   & 0.333 & 1 & 1 & 20 & 0 & 0\\
     miniBELEN-12  & 0.467 & 0.933 & 1 & 16 & 2 &0\\
    \botrule
    \end{tabular*}
\end{table}

\begin{figure}[!ht]
    \centering
    \includegraphics[width=1\textwidth]{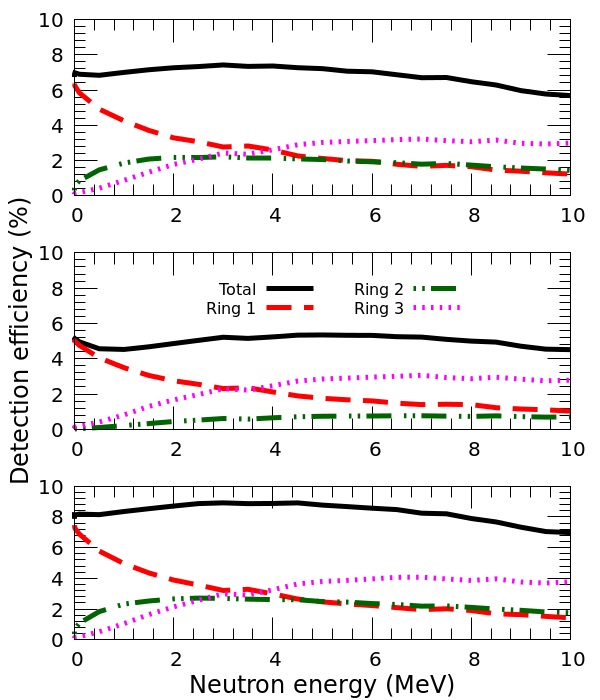}
    \caption{(From top to down) Calculated neutron detection efficiency of miniBELEN-10A, miniBELEN-10B and miniBELEN-12. The y-axis error-bars are smaller than the size of the data points. Black solid: total efficiency. Red dashed: ring 1 (inner) efficiency. Green dash-dotted: ring 2 efficiency. Magenta dotted: ring 3 (outer) efficiency.}
    \label{fig:confDesigns_Plots}
\end{figure}

Figure \ref{fig:confDesigns_Plots} shows the Monte Carlo-calculated efficiencies of each configuration of the miniBELEN detector. Calculations include the implementation of the composition functions using the cadmium filters described in figure \ref{fig:cdfilters}. In miniBELEN-10A and miniBELEN-12 configurations the counters are distributed using the same structure of rings so that they present a similar response. However, because of the use of two extra counters in the outer ring, the total efficiency is larger in miniBELEN-12. On the other hand, miniBELEN-10B presents a different structure of rings (the outer and the middle rings could be actually merged in a single outer pseudo-ring) and, consequently, a different response. As shown in table \ref{tab:effFlat}, in miniBELEN-10B the flatness up to 10 MeV is better than in the other configurations but at the cost of worsening at the flatness up to 5 MeV.

\begin{table}[h]
    \caption{Average efficiency up to 10 MeV ($\varepsilon$) and flatness values ($F$) up to 5 MeV, 8 MeV and 10 MeV for each configuration of miniBELEN (see text for more details).}\label{tab:effFlat}
    \begin{tabular*}{\textwidth}{@{\extracolsep\fill}lcccc}
    \toprule%
    Configuration & $\varepsilon_{av} (10)$ & $F (5)$ & $F (8)$ & $F (10)$ \\
    \midrule
     miniBELEN-10A & 6.90(3)\% & 1.102(6) & 1.146(7)&  1.306(7) \\
     miniBELEN-10B   & 5.00(2)\% & 1.181(8) &1.181(8) & 1.184(8)  \\
     miniBELEN-12  & 8.26(3)\%& 1.132(6)  & 1.132(6)& 1.277(7) \\
    \botrule
    \end{tabular*}
\end{table}

\newpage

\subsection{Discussion}
\begin{figure}[!ht]
    \centering
    \includegraphics[width=1\textwidth]{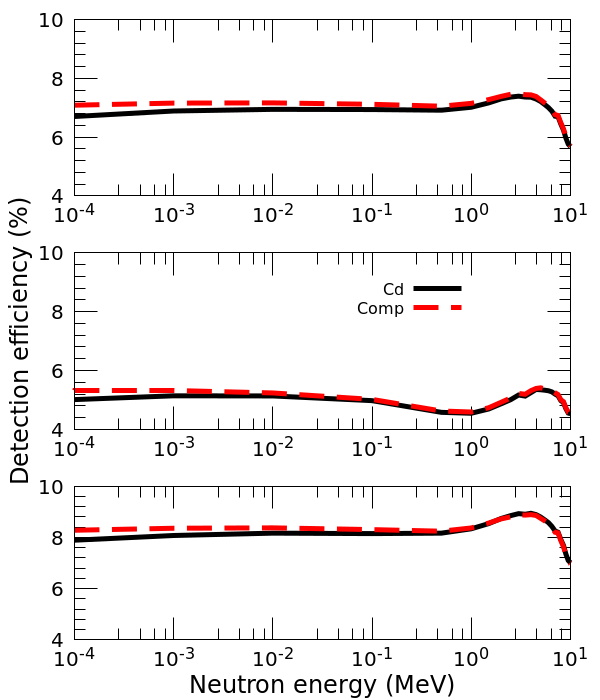}
    \caption{(From top to down) Calculated neutron detection efficiencies of miniBELEN-10A, miniBELEN-10B and miniBELEN-12. Black solid: results when implementing the composition functions using cadmium filters. Red dashed: results when the composition functions are implemented numerically.}
    \label{fig:CdEffectsPlot}
\end{figure}

The design methodology was based on the assumption that the effect of the composition function could be physically implemented using neutron filters. In particular, miniBELEN uses thick cadmium filters so that all neutrons below 0.5 eV (the so-called cadmium cut-off) are absorbed. However, some neutrons above the cadmium cut-off could still be detected in the $^3$He counters. This results in an overestimation of the detection efficiency at low neutron energies (see figure \ref{fig:CdEffectsPlot}) during the design process. However, the overestimation does not significantly affect the values of average efficiency and the flatness parameter.

A summary of the main characteristics (average efficiency and flatness) of the moderated neutron counters designed in recent decades is provided in table \ref{tab:detectors}. Most of them present flat efficiencies up to approximately 1 MeV, which was suitable for $\beta$-delayed neutron measurements (the original application of these detection systems) but not for most of the ($\alpha,n$) reactions. In recent years, efforts have been made to achieve moderated neutron counters with flat efficiencies above 1 MeV.

\begin{table}[h]
    \caption{Comparison of of the average efficiency $\varepsilon_{av}$ and flatness $F$ for various moderated neutron counters designed in the past decades. See section \ref{sec:secParMer} for more details.}\label{tab:detectors}
    \begin{tabular*}{\textwidth}{@{\extracolsep\fill}lcccccc}
    \toprule%
    Detector & $F$(1 MeV) & $\varepsilon_{av}$(1 MeV) & $F$(5 MeV) & $\varepsilon_{av}$(5 MeV)\\
    \midrule
    NERO\footnotemark[0] \cite{PEREIRA2010275} & 1.157 & 43.1\% & 1.926 & 38.3\%\\
    LOENIE\footnotemark[0] \cite{LMathieu_2012} & 1.016 & 17.4\% & 1.645 & 16.1\% \\
    MAINZ \cite{LMathieu_2012} & 1.130 & 47.0\% & 2.245 & 39.9\%\\
    Falahat \textit{et al.} \cite{falahat20133he} & 1.197 & 55.2\% & \textit{No data} & \textit{No data} \\
    Laurent \textit{et al.} \cite{LAURENT201499} & 1.080 & 35.9\% & 1.739 & 32.5\% \\
    Hybrid-3Hen\footnotemark[0] \cite{grzywacz2014hybrid} & 1.103 & 36.8\% & 1.781 & 32.5\% \\
    TETRA\footnotemark[0] \cite{TESTOV201696} & 1.133 & 61.1\% & 1.842 & 51.6\% \\
    BRIKEN\footnotemark[1] \cite{tarifeno2017conceptual} & 1.021 & 66.3\% & 1.263 & 60.9\%\\
    \toprule%
    Detector & $F$(5 MeV) & $\varepsilon_{av}$(5 MeV) & $F$(8 MeV) & $\varepsilon_{av}$(8 MeV)\\
    \midrule
    FED\footnotemark[2]\cite{utsunomiya2017direct} & 1.164 & 36.37\% & 1.368 & 34.51\%\\
    HeBGB \cite{Brandenburg_2022} & 1.170 & 7.41\%  & 1.179 & 7.57\%\\
    ELIGANT-TN \cite{clisu2023cross} & 1.131 & 36.72\% & 1.343 & 34.57\%\\
    \botrule
    \end{tabular*}
    \footnotetext[0]{Data from reference \cite{tarifeno2017conceptual}.}
    \footnotetext[1]{Neutron response privately communicated by co-author M. Pallàs.}
    \footnotetext[2]{Neutron response extrapolated for energies higher than 6 MeV.}
\end{table}

\clearpage

\newpage
\section{Implementation of miniBELEN-10A}\label{sec:Implementation}
As it will be discussed in detail in section \ref{sec:comissioning}, it was decided to carry out the commissioning of miniBELEN using the relatively well-known $^{27}$Al($\alpha,n$)$^{30}$P reaction. Since the reaction threshold is 3014.81(9) keV \cite{qcalc} and the alpha energies involved in the commissioning experiment were not greater than 10 MeV, the maximum energy of the produced neutrons was not expected to exceed 7 MeV. From table \ref{tab:effFlat}, it can be concluded that miniBELEN-10A and miniBELEN-12 are more suitable (in terms of efficiency flatness) for carrying out this measurement than miniBELEN-10B. It was finally decided to implement miniBELEN-10A due to hardware limitations (only ten neutron counters were available). However, the only relevant difference between miniBELEN-10A and miniBELEN-12 is the magnitude of the efficiency.

\subsection{Physical implementation of the detector}
miniBELEN-10A uses ten $^{3}$He-filled neutron proportional counters which are embedded in the modular block of HDPE as shown in figure \ref{fig:confDesigns}. Cadmium filters, as has been described in the previous section, are used in order to achieve a detection efficiency that is nearly independent from the initial neutron energy (see figure \ref{fig:cdfilters} and table \ref{tab:compoFunVal} for more details). In order to reduce the contribution of the room background (scattered neutrons coming from all directions), the laterals of the detector were covered with thick cadmium foils, while the upper and lower faces were covered with a thick layer of boron-silicone. Additionally, a shielding made up of HDPE and boron-silicone is used to attenuate the in-beam background component (neutrons produced by the interaction of the alpha particles with the elements of the beam line and directly coming from this direction).

Two versions of miniBELEN-10A have already been implemented. Both are based on the use of the $^3$He-filled proportional counters manufactured by LND which are described in section \ref{sec:detDesing}. The only difference is the pressure of the gas. The first version (since now, v2021) was implemented using readily available hardware as a demonstrator prototype. The specifications of the $^{3}$He-filled counters used in v2021 are listed in table \ref{tab:tubemodels1}. In the definitive implementation of the detector (since now, v2022), the gas pressure of all counters was 8 atm (LND model 252285). 

\begin{table}[h]
    \caption{Gas pressure of the LND \cite{LND} neutron counters used in miniBELEN-10A-v2021. $N_i$ is the number of counters used in ring $i$. See section \ref{sec:detDesing} for more details.}\label{tab:tubemodels1}
    \begin{tabular*}{\textwidth}{@{\extracolsep\fill}lcccc}
    \toprule%
    LND model & Gas pressure & $N_1$ & $N_2$ & $N_3$ \\
    \midrule
     252248 & 20 atm & 0 & 0 &  1 \\
     252266 & 10 atm & 2 & 2 & 3 \\
     252285 & 8 atm & 0 & 0 & 1 \\
     252303 & 4 atm & 0 & 0 & 1\\
    \botrule
    \end{tabular*}
\end{table}

\newpage
\begin{figure}[!ht]
    \centering
    \includegraphics[width=0.9\textwidth]{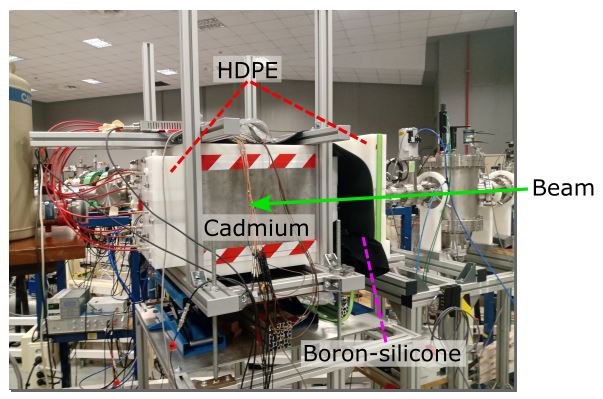}
    \caption{Definitive implementation of miniBELEN-10A at CMAM. The red dashed line points at the high-density polyethylene (HDPE) in miniBELEN (left) and in the in-beam shielding (right). The pink dashed line points at the boron silicone used for the shielding in-beam. The green arrow shows the direction of the alpha-particles beam. The grey plate is the the cadmium shielding.}
    \label{fig:SetupDefMB}
\end{figure}

\subsection{Neutron detection efficiency by Monte Carlo simulations} \label{sec:effDef}
\begin{figure}[!ht]
    \centering
    \includegraphics[width=1\textwidth]{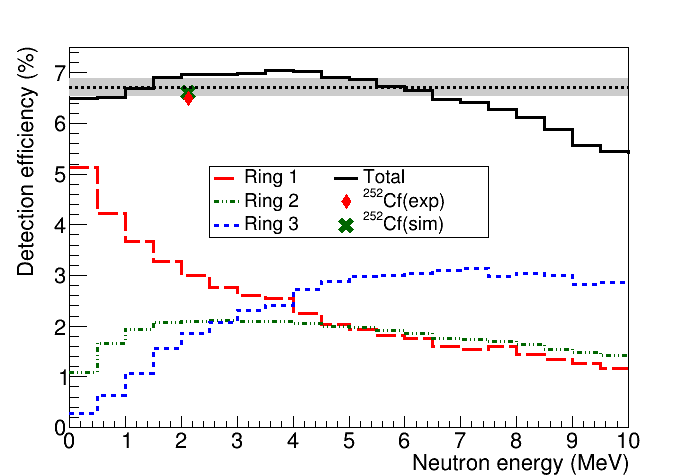}
    \caption{Response of miniBELEN-10A (v2022) by Monte Carlo calculations. Black solid line: total efficiency. Red dashed line: ring 1. Green dotted-dashed line: ring 2. Blue dotted line: ring 3. The colored band is centered at the nominal value of the efficiency (black dotted line) and shows the systematic uncertainty due to the flat efficiency hypothesis. The simulated (green cross) and experimental (red rhombus) efficiencies for a $^{252}$Cf source are also shown (see section \ref{sec:cf252char}).}
    \label{fig:FinalEff}
\end{figure}

Monte Carlo calculations using \textit{Particle Counter} were carried out in order to determine the neutron detection efficiency of miniBELEN-10A. The efficiency was computed as a response function. Calculations were carried out by assuming a point-like isotropic neutron source placed at the center of the detector. The results are shown in figure \ref{fig:FinalEff}.

Since miniBELEN-10A was designed to have a flat efficiency, a sensitivity study was performed to assess how different ($\alpha$,n) energy spectra affect the total detection efficiency. To do this, a set of ($\alpha$,n) spectra are generated using SaG4N \cite{MENDOZA2020163659}, a Monte Carlo tool based on Geant4. For each spectra, a convoluted efficiency is calculated as,

\begin{equation}\label{eq:convEfficiency}
  \varepsilon^i = \sum_{k = 0}\varepsilon_k S_k^i
\end{equation}

\begin{figure}[!ht]
    \centering
    \includegraphics[width=1\textwidth]{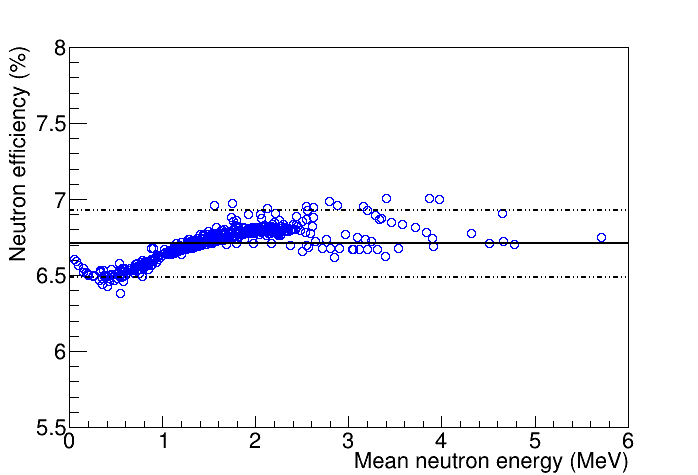}
    \caption{Scatter plot showing the convoluted efficiency as function of the mean neutron energy. The solid black line is the average value of the efficiency distribution (nominal efficiency) while the dashed lines shows the standard deviation with 95\% confidence (see text for more details).}
    \label{fig:EffFlat_Daniel}
\end{figure}

$S_k^i$ are the bins of the $i$-th spectra, and $\varepsilon_k$ are the efficiency bins. The distribution of convoluted efficiencies is then constructed as a function of the mean neutron energy (see figure \ref{fig:EffFlat_Daniel}). For that purpose, a set of ($\alpha$,n) spectra calculated for the most common isotope of each stable element with 3 $<$ Z $<$ 82 was used. Calculations have been carried out using alpha energies up to 15 MeV, which is the maximum available beam energy at CMAM (at CNA is 9 MeV). It is interesting to note that for all the spectra studied the average neutron energy is always below 6 MeV. 

\clearpage

The nominal efficiency $\varepsilon$ of miniBELEN is computed as the average value of the distribution of convoluted efficiencies (solid black line in figure \ref{fig:EffFlat_Daniel}) while the systematic uncertainty introduced by the flat efficiency hypothesis (color band in figure \ref{fig:FinalEff} and dashed lines in \ref{fig:EffFlat_Daniel}) is taken as the standard deviation with 95\% confidence. Then,

\begin{equation}
   \varepsilon = 6.71(18)\%
\end{equation}

\clearpage

\subsection{Experimental characterization with a fission source}  \label{sec:cf252char}
\subsubsection{Neutron Multiplicity Counting (NMC)}
The Monte Carlo simulations of the detection efficiency were validated using a $^{252}$Cf fission source and the Neutron Multiplicity Counting (NMC) technique. The method is based on the use of the emission of correlated neutrons in spontaneous fission processes and was originally aimed at characterizing materials, especially $^{240}$Pu, for nuclear safeguard applications \cite{osti_1679}. More recently, the NMC technique has been shown to be suitable for a high-precision determination of $^{252}$Cf neutron emission yields \cite{Henzlova_2019,pallas2022efficiency}.

For a point-like source where the neutron emission trough ($\alpha, n$) reactions and induced fission processes is negligible, the so-called Singles ($S$, i.e. the total neutron counting rate) and Doubles ($D$, i.e. the counting rate of correlated neutron pairs) can be expressed as \cite{croft_etal_2013}, 
\begin{equation}
  \label{eq:NMCS}
  S = F\varepsilon(\nu_1 + \nu_d)
\end{equation}
\begin{equation}
  \label{eq:NMCD}
  D = \frac{F\varepsilon^2\nu_2}{2}
\end{equation}

Where $F$ is the spontaneous fission rate, $\varepsilon$ is the total neutron detection efficiency, $\nu_1$ is the mean number of prompt neutrons emitted per fission (that is, the first moment of the neutron multiplicity distribution), $\nu_d$ is the mean number of delayed neutrons and $\nu_2$ is the second moment of the emitted neutron multiplicity distribution. Consequently, the neutron detection efficiency can be determined from the measured Singles and Doubles as,

\begin{equation}
  \label{eq:NMCeff}
  \varepsilon = 2\frac{\nu_1 + \nu_d}{\nu_2}\frac{S}{D} = N\frac{S}{D} 
\end{equation}

Where $N$ = 0.631(2) \cite{croft_etal_2013} for $^{252}$Cf.

\subsubsection{Experimental setup}\label{sec:setup}
Neutrons emitted from the $^{252}$Cf source were moderated by the HDPE and detected in the $^3$He-filled counters. Each counter was connected to a charge-sensitive preamplifier (Canberra 2006 and Mesytech MPRS-16 modules). The output signals were sent to the acquisition system (DAQ), which consisted of a SIS3316 sample digitizer (16 channels and 250 MHz sampling frequency) from Struck Innovative Systeme controlled by the \textit{Gasific 7.0} software \cite{AGRAMUNT201669}. The result is file containing a list of the events registered in each counter with information of the pulse amplitude (i.e., energy deposited) and a time stamp (list mode data). Correlations between events are analyzed offline in order to extract the Singles and the Doubles.

The neutron source was a glass vial containing 1 mL of a 0.1 M californium solution in aqueous nitric acid. The geometry of the vial can approximated as a 0.66 cm radius and 4.5 cm length cylinder. The vial was placed at the center of the detector. The isotopic composition (in activity, measured on 13/03/2019) of the radioactive source was 83.089\% $^{252}$Cf, 16.323\% $^{250}$Cf, 0.4881\% $^{249}$Cf and 0.0991\% $^{251}$Cf. The decay product $^{248}$Cm from $^{252}$Cf was last separated on 22/08/2014.

\subsubsection{Results}
Of all the isotopes in the californium source, only $^{252}$Cf, $^{250}$Cf are relevant neutron emitters. From equation \ref{eq:NMCeff}, using data from \cite{osti_1679} and \cite{Henzlova_2019} and information about the isotopic composition provided by the manufacturer, it is found that for our neutron source $N$ = 0.642(2). Therefore, the measured Singles and Doubles were used to determine the neutron detection efficiency of miniBELEN-10A. This value can be compared with a Monte Carlo simulation for a pure $^{252}$Cf source (the neutron energy spectrum from reference \cite{radevCf} was used). The results are shown in table \ref{tab:NMCresults}. A relative overestimation of the experimental efficiency not greater than 2\% is found. See (figure \ref{fig:FinalEff}) that for both the experiment and the simulation, the results are consistent with the nominal value of the efficiency obtained from the flat response hypothesis ($\varepsilon$).

 \begin{table}[h]
    \caption{Experimental (NMC) and simulated (MC) efficiencies for $^{252}$Cf in miniBELEN-10A. The nominal efficiency ($\varepsilon$) coming from the flat response hypothesis is also shown. The uncertainty of the Monte Carlo efficiency is purely statistical.}\label{tab:NMCresults}
    \begin{tabular*}{0.9\textwidth}{@{\extracolsep\fill}cccc}
    \toprule%
    NMC efficiency & MC efficiency & NMC/MC & $\varepsilon$ nominal\\
    \midrule
        6.50(3)\% & 6.615(7)\% & 0.983(4) & 6.71(18)\%\\
    \botrule
    \end{tabular*}
\end{table}

\subsubsection{Magnitudes affecting the efficiency}\label{sec:magnitudes}
Monte Carlo simulations overestimate the experimental efficiency (see table \ref{tab:NMCresults}). This is not a surprising result since in such type of detectors Monte Carlo simulations always tend to overestimate the detection efficiency \cite{falahat20133he,PEREIRA2010275,brandenburg2022measurements} due to a combination of uncertainties in the neutron scattering cross sections, the density of the moderator and the pressure of the neutron proportional counters. In order to account for these discrepancies, the effect of modifying the last two magnitudes has been studied.

The range of variation for the density of the moderator was established after an experimental characterization of the polyethylene. For the gas pressure, the manufacturer does not declare any uncertainty. However, it is possible to establish a realistic range of variation knowing that the counters include a small amount of CO$_2$ (quenching gas) of about 1 to 3\%. For each magnitude studied, $\Delta \varepsilon$ is defined as the difference between the maximum and minimum values of the efficiency within the range of variation. 

\begin{table}[h]
    \caption{Effects of the HDPE density ($\rho$) and the $^{3}$He pressure ($P$) on the neutron efficiency for a $^{252}$Cf source. See main text for more details.}\label{tab:magEff}
    \begin{tabular*}{\textwidth}{@{\extracolsep\fill}ccccc }
    \toprule%
        Magnitude & Range of & Efficiency  & Difference & NMC efficiency \\
         & variation & variation & ($\Delta \varepsilon$) & (see table \ref{tab:NMCresults})\\
    \midrule
        $\rho$ (g/cm$^3$) & 0.947 to 0.951 & 6.598(7) to 6.632(7)\% & 0.034(10)\% & 6.50(3)\%\\
        $P$ (atm) & 7.76 to 8.00 & 6.557(7) to 6.615(7)\% & 0.058(10)\% & 6.50(3)\%\\
    \botrule
    \end{tabular*}
\end{table}

From table \ref{tab:magEff}, we can first conclude that the gas pressure could explain the small discrepancy with the $^{252}$Cf measurement. Moreover, it is interesting to note that the systematic uncertainty of the flat efficiency hypothesis encompasses the uncertainty that can be attributed to both the gas pressure and the density of the moderator. Summing up the three contributions we obtain that,

\begin{equation}
    \label{eq_UncertaintyEff_Final}
    \delta \varepsilon = \sqrt{(\delta \varepsilon_{f})^2+(\delta \varepsilon_{\rho})^2+(\delta \varepsilon_{p})^2} = 0.18\%
\end{equation}

Where $\delta \varepsilon_{f}$ is the contribution of the flat efficiency hypothesis, $\delta \varepsilon_{\rho}$ is the systematics due to the uncertainty in the HDPE density and $\delta \varepsilon_{P}$ is the systematics from the uncertainty in the gas pressure. The values of $\delta \varepsilon_{\rho}$ and $\delta \varepsilon_{p}$ come from $\Delta \varepsilon$ in table \ref{tab:magEff}.

\subsection{Effect of the source position}\label{sec:sourceCorrections}
\begin{figure}[!ht]
    \centering
    \includegraphics[width=0.8\textwidth]{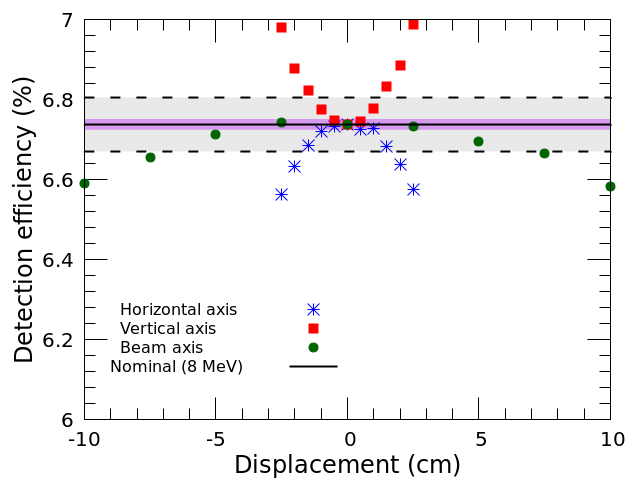}
    \caption{Efficiency of miniBELEN-10A as function of the source position, shown as a displacement in respect of the central position (0 cm) along the different axis. Green circles: beam axis. Red squares: vertical axis. Blue crosses: horizontal axis. Black line: nominal efficiency up to 8 MeV. The colored gray-band between dashed lines limits the region where relative variations of the efficiency are smaller than 1\% while the violet region shows the range of positions where these relative variations are smaller than 0.2\%. The error-bars are smaller than the size of the data-points.}
    \label{fig:positions}
\end{figure}

The nominal efficiency of miniBELEN was computed for a point-like and isotropic located at the center of the detector. The effects of the source positioning have been studied later. Figure \ref{fig:positions} shows the detection efficiency for a point-like $^{252}$Cf source at different positions along the beam axis as well as the horizontal and vertical axis, which are perpendicular to the former. In all graphs, the gray region shows the range of positions where relative variations in efficiency with respect to the central position are not greater than 1\%, while the violet region shows the range of positions where these relative variations are smaller than 0.2\%. When a measurement using a point-like source is performed, the source should be positioned within these margins to ensure that the nominal efficiency value can be used.

\newpage
\section{Commissioning for ($\alpha$,n) reactions}\label{sec:comissioning}
The commissioning of miniBELEN-10A was carried out by measuring the $^{27}$Al($\alpha$,n)$^{30}$P thick-target yields at the 45º beam-line in Centro de Micro-Análisis de Materiales (CMAM, Madrid, Spain) \cite{CMAM,redondo2021current}. There are various reasons that justify why the reaction was selected for the commissioning measurement. First, the thick-target yields are well known for alpha energies near the reaction threshold (3.014 MeV \cite{qcalc}) up to 10 MeV. Secondly, evaluated data on reaction cross section and production yields \cite{murataJENDL,vlaskin2015neutron} exist. Last but not least, it is easy (and not so expensive) to obtain a high purity target of $^{27}$Al.

\subsection{Experimental setup}
The alpha-particles beam (charge state +2e) was accelerated using the 5 MV tandem accelerator at CMAM and passed through a 7 mm diameter tantalum ($^{181}$Ta) collimator. The target consisted on a high-purity natural aluminum ($^{27}$Al) thick foil which was placed close to the collimator (14 mm) and in the center of the detector. Note that the diameter of the collimator ensures that the "neutron source" (i. e., the region where $\alpha$-induced reactions take place) is small enough to be considered point-like (see section \ref{sec:magnitudes} for more details). 

The beam current was continuously monitored online by a custom readout and suppression device and an ORTEC-439 Digital Current Integrator which produced an output pulse for each 10$^{-10}$ C collected at the target. The output pulses were sent to the same data acquisition system used to process the signals from the $^{3}$He-filled counters. A +300 V bias voltage was applied to the collimator in order to prevent secondary electrons from escape the target.

\subsection{Determination of the thick target yields}
The thick-target neutron production yield $Y(E_\alpha)$ at a given alpha particle energy $E_\alpha$ was determined as follows,

\begin{equation}
  \label{eq:yields}
  Y(E_\alpha) = \frac{\sum_{i = 1}^{10} R_i}{\varepsilon I} = \frac{R_T}{\varepsilon} = \frac{r}{\varepsilon}
\end{equation}

Where $R_i$ is the neutron rate in counter $i$ ($R_T$ is the total detector rate), $\varepsilon$ is the nominal neutron detection efficiency calculated as described in Section \ref{sec:effDef}, $I$ is the number of alpha particles reaching the target per time unit (that is, the beam current), and $r$ is the quotient between the total rate and the beam current, also called the current normalized rate.

\subsection{Neutron background}
A detailed characterization of the neutron background showed that it could be broken down into three different components. The first one was the room background, which were all neutrons not produced by the accelerator facility and were found to be completely negligible.  The second component was the in-beam background, which was produced by the interaction of the alpha-particles with the elements of the beam line. In-beam shielding (see figure \ref{fig:SetupDefMB}) was specifically designed to mitigate such a background. Measurements showed that the use of in-beam shielding reduced the neutron background by a relative factor 15\%. The third and most relevant component was due to the impurities in the region of the target holder and possibly in-beam neutrons not absorbed by the shielding. This was consistent with the fact that the background rates were larger in the counters of the inner ring.

The background component measurements were performed using a dummy target consisting of a thick natural tantalum ($^{181}$Ta) foil. It should be kept in mind that the ($\alpha,n$) threshold for $^{181}$Ta is 10.009(5) MeV \cite{qcalc}. 

\subsection{Results of the commissioning}
\begin{figure}[!ht]
    \centering
    \includegraphics[width=0.8\textwidth]{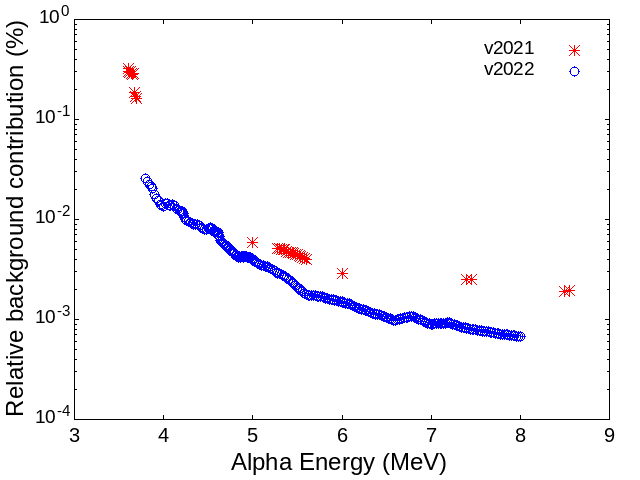}
    \caption{Relative background contribution measured using the prototype version (v2021, red crosses) of miniBELEN and the definitive implementation of the detector (v2022, blue circles).}
    \label{fig:bckRatios}
\end{figure}

Measurements have been carried out using both versions of miniBELEN, the prototype demonstrator (v2021) and the definitive implementation of the detector (v2022). Figure \ref{fig:bckRatios} shows the contribution of the neutron background at different beam energies. At low energies (3 - 4 MeV), the contribution is clearly not negligible. Above 5 MeV it is of the same order as the statistical relative uncertainties, which range from about 0.5 - 1\% (low energy) to 0.03\% (at 8.5 MeV).

\begin{figure}[!ht]
    \centering
    \includegraphics[width=0.8\textwidth]{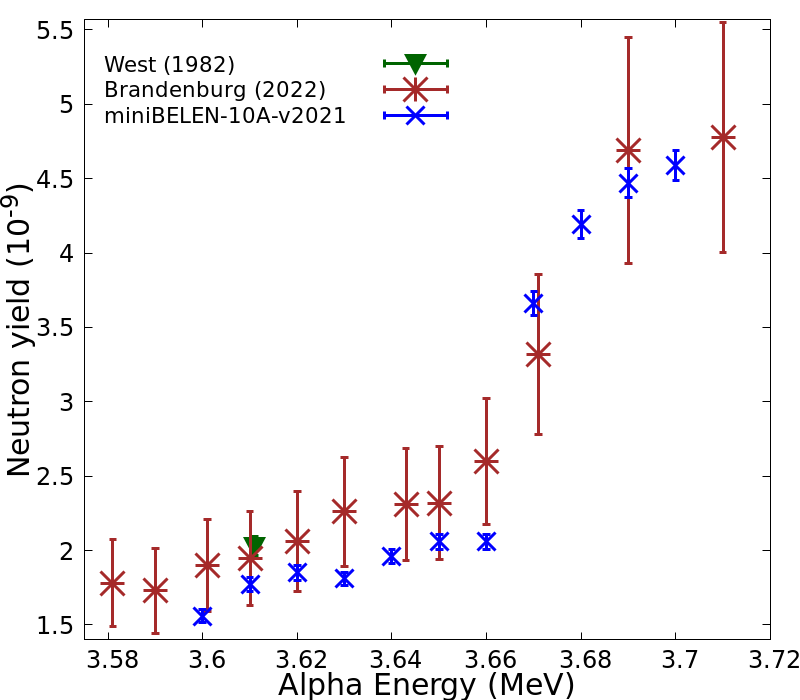}
    \includegraphics[width=0.8\textwidth]{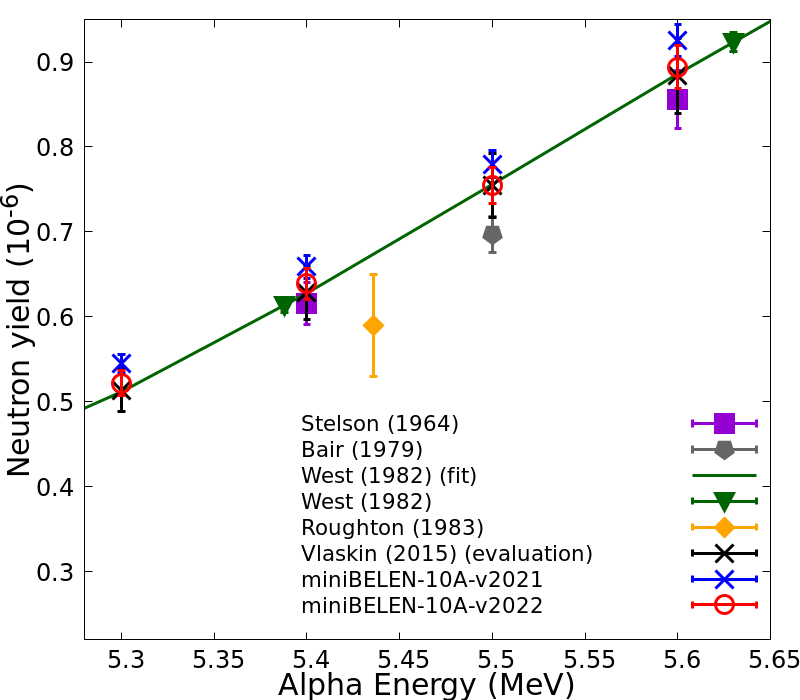}
    \caption{$^{27}$Al($\alpha,n$)$^{30}$P thick-target yields (see main text for more information). The green line is a cubic spline fit in the data from West \textit{et al.} \cite{west1982measurements}.}
    \label{fig:AlYields}
\end{figure}

\begin{figure}[!ht]
    \centering
    \includegraphics[width=0.8\textwidth]{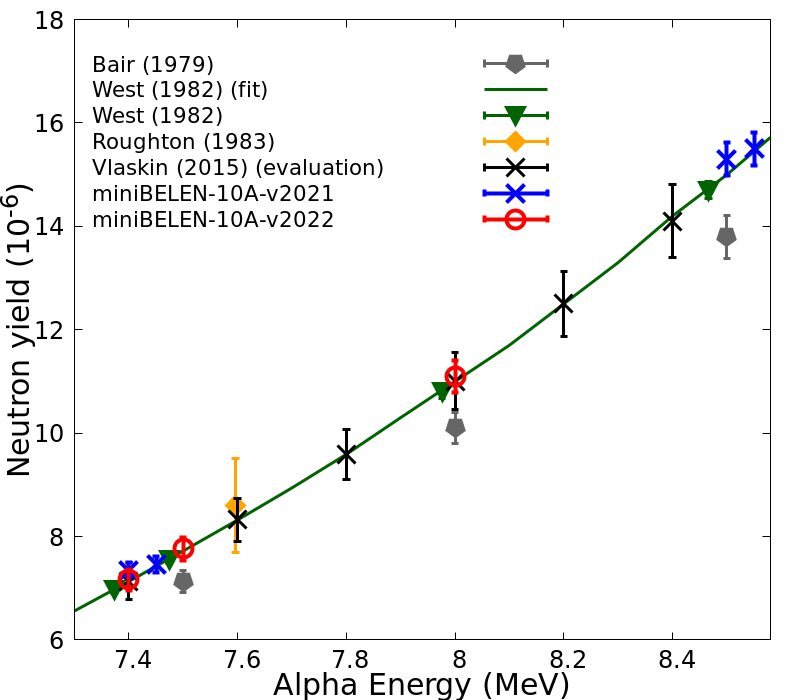}
    \caption{$^{27}$Al($\alpha,n$)$^{30}$P thick-target yields (see main text for more information). The green line is a cubic spline fit in the data from West \textit{et al.} \cite{west1982measurements}.}
    \label{fig:AlYields3}
\end{figure}

The dead-time and background corrected yields from $^{27}$Al($\alpha,n$)$^{30}$P measured with miniBELEN are shown in figures \ref{fig:AlYields} - \ref{fig:AlYields3}. For the counting rates involved in this experiment (less than 10 k counts per counter), the systematics introduced by the dead-time correction method is negligible. Thus, the dominant contribution to the total uncertainty is the systematics introduced by the flat-efficiency hypothesis.

In figures \ref{fig:AlYields} and \ref{fig:AlYields3} the neutron yields measured with both versions of miniBELEN-10A (the v2021 demonstrator and the v2022 definitive implementation) are compared with previously existing data obtained by direct neutron counting (Stelson \cite{stelson1964cross}, Bair \cite{bair1979neutron}, West \cite{west1982measurements} and Brandenburg \cite{brandenburg2022measurements}), from activation measurements (Roughton \cite{ROUGHTON1983341}) and from energy spectra measured using the time-of-flight technique (Jacobs \cite{JACOBS1983541}). A recent evaluation by Vlaskin \textit{et al.} \cite{vlaskin2015neutron} is also included.

\clearpage

\clearpage

\subsection{Determination of the mean neutron energy }
The design of miniBELEN implies that it is not possible to know the initial energy of the neutrons detected in the $^3$He counters. However, it is possible to have some knowledge about the neutron energy distribution, in particular, the mean value of the distribution. The idea is to use the energy dependence of the ring responses in order to correlate the counting rate ratios with the neutron energies \cite{HeederPhysRevC.15.2098,Heeder01091981_1981}.

Let $r_i$ be the current normalized rate (see equation \ref{eq:yields}) in ring $i$ and $\varepsilon_i$ be the detection efficiency in that ring. The quantities $\hat{\varepsilon}$ and $\hat{r}$ should be equivalent under the following definitions.

\begin{equation}
    \hat{\varepsilon} = \frac{\varepsilon_1}{\varepsilon_1+\varepsilon_2+\varepsilon_3} + \frac{\varepsilon_2}{\varepsilon_3}
\end{equation}
    
\begin{equation}
    \hat{r} = \frac{r_1}{r_1+r_2+r_3} + \frac{r_2}{r_3}
\end{equation}

\begin{figure}[!ht]
    \centering
    \includegraphics[width=1\textwidth]{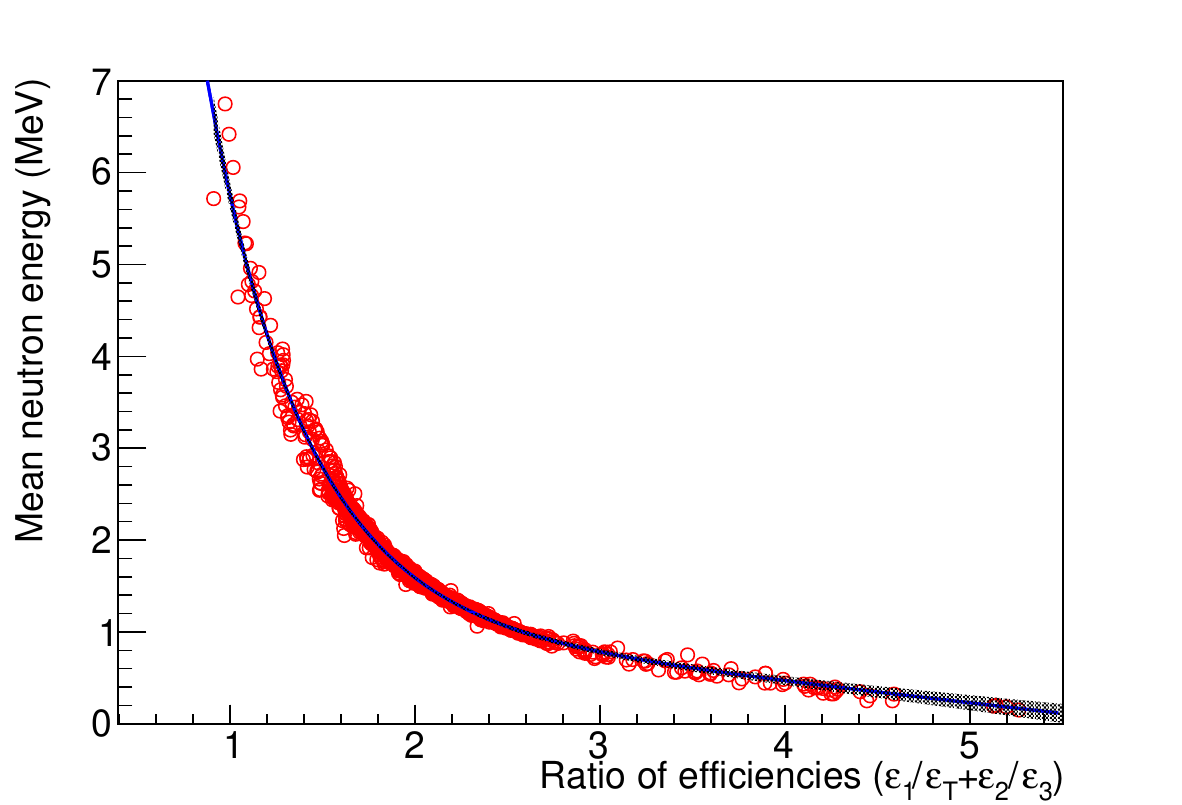}
    \caption{Dependence of the mean energy with the rings efficiency ratio $\varepsilon_1/\varepsilon_T + \varepsilon_2/\varepsilon_3$, with $\varepsilon_T$ the total detection efficiency. The ($\alpha$,n) spectra described in section \ref{sec:effDef} have been used to obtain the convoluted efficiency of each ring. The blue line is the calibration curve $\hat{E} = \exp(a_0+a_1\hat{\varepsilon})+a_2\hat{\varepsilon}+a_3$ while the black colored band show the 99\% confidence intervals of the fit.}
    \label{fig:MeanEnRatio}
\end{figure}

The mean neutron energy $\hat{E}$ is therefore extracted from the calibration curve as illustrated in figure \ref{fig:MeanEnRatio}. The analytical expression $\hat{E} = \exp(a_0+a_1\hat{\varepsilon})+a_2\hat{\varepsilon}+a_3$ was fitted to the data points obtained from the Monte Carlo calculated ($\alpha$,n) spectra described in section \ref{sec:effDef}. The value of the fit parameters is shown in table \ref{tab:MeanEffTable}.

\clearpage

In order to test the consistency of the methodology for extracting the mean neutron energy, the experimental results with miniBELEN are compared with the $^{27}$Al($\alpha$,n)$^{30}$P neutron energy spectra measured by Jacobs \textit{et al.} \cite{JACOBS1983541}. The results are shown in table \ref{tab:MeanEffTable}. The same exercise can be performed for a $^{252}$Cf source. When the miniBELEN-extracted mean energies are compared with data from experimental spectra, the results for the $^{252}$Cf source are in better agreement. It must be kept in mind that for the $^{27}$Al we rely on a single experimental measurement (the one by Jacobs \textit{et al.})


\begin{table}[h]
    \caption{$^{27}$Al($\alpha$,n)$^{30}$P mean neutron energies (in MeV) measured with miniBELEN compared with the values obtained from experimental spectra. Results for a $^{252}$Cf fission source are also included.}\label{tab:MeanEffTable}
    \begin{tabular*}{\textwidth}{@{\extracolsep\fill}ccc }
    \toprule%
        Beam energy & miniBELEN & Experimental spectra  \\
    \midrule
        4.0 MeV & 0.59(10) & 0.515 \cite{JACOBS1983541}\\
        4.5 MeV & 0.78(8) & 0.676 \cite{JACOBS1983541}\\
        5.0 MeV & 1.04(9) & 0.904 \cite{JACOBS1983541}\\
        5.5 MeV & 1.24(7) & 1.079 \cite{JACOBS1983541}\\
        $^{252}$Cf source & 2.00(7) & 2.125 \cite{BROWN20181ENDFVIII}\\
    \botrule
    \end{tabular*}
\end{table}

\clearpage
\section{Concluding remarks}\label{sec:Conclusions}
miniBELEN is a moderated neutron counter consisting on an array of $^{3}$He-filled proportional counters that are embedded in a modular block of high-density polyethylene. The modular structure makes possible to obtain different types of response by simply arranging the same material in different ways (different configurations). An innovative design methodology (the so-called composition method) based on the use of (cadmium) neutron filters has been used to achieve a detection which is nearly independent from the initial neutron energy (flat efficiency). 

The miniBELEN-10A configuration has been physically implemented and experimentally characterized using a $^{252}$Cf fission source. The results show that Monte Carlo simulations tend to overestimate the neutron efficiency, with a relative discrepancy not greater than 2\%, which is compatible with the declared systematic uncertainty due to the flat efficiency hypothesis. The effects on the total efficiency of the $^{3}$He pressure, the HDPE density, and the positioning of the source have been evaluated and found to be negligible compared to the nominal systematics.

Finally, the commissioning of the detector has been carried out by measuring the $^{27}$Al($\alpha,n$)$^{30}$P thick-target yields at Centro de Micro-Análisis de Materiales (CMAM). Two versions of miniBELEN-10A, a prototype and the final implementation, have been used. The results of both measurements are consistent with previously existing experimental data and evaluations. Moreover, a method for determining the mean neutron energy from the counting rate ratios has been developed. 

\bmhead{Acknowledgments}
This work has been supported by the Spanish government grants FPA2017-83946-C2-1 and C2-2, PID2019-104714GB-C21 and C22, PID2021-126998OB-I00 and RTI2018-098868-B-I00, the Generalitat Valenciana grant PROMETEO/2019/007, the SANDA project funded under the H2020 EURATOM-1.1 grant 847552, and the MRR (Recuperation and Resilience Mechanism) cofinanced by the Generalitat de Catalunya, the Spanish Ministerio de Ciencia e Investigación (MCIN) and the European Union (EU).  The authors acknowledge the support from Centro de Micro-Análisis de Materiales (CMAM) - Universidad Autónoma de Madrid (UAM) for the beam time proposals with codes P01075, P01156 and P01302 and to its technical staff for their contribution to the operation. The author Nil Mont-Geli also acknowledges Banco de Santander for its funding and the grant 2021 FISDU 00368 financed by the Secretaria d'Universitats i Recerca del Departament d’Empresa i Coneixement de la Generalitat de Catalunya.

\bmhead{Data availability statement}

Data can be made available by the corresponding author upon reasonable request. 

\clearpage
\bibliography{sn-bibliography}

\end{document}